\documentclass[epsfig,12pt]{article}

\usepackage{latexsym}
\usepackage{relsize}
\usepackage{geometry}
 \usepackage{amsmath, amssymb, amscd, xypic, graphicx}
\usepackage{slashed,color}
\usepackage{epstopdf}

\usepackage[colorlinks]{hyperref}
\usepackage{amsmath, mathrsfs}
\usepackage[displaymath, tightpage]{preview}

\input xy
\xyoption{all}

 \usepackage{fancyhdr}                   
\def\beq{\begin{equation}}
\def\eeq{\end{equation}}
\def\beqn{\begin{eqnarray}}
\def\eeqn{\end{eqnarray}}

\newcommand{\sma}{\left(\begin{smallmatrix}}
\newcommand{\smaa}{\end{smallmatrix}\right)}

\newcommand{\gsim}{\lower.7ex\hbox{$
\;\stackrel{\textstyle>}{\sim}\;$}}
\newcommand{\lsim}{\lower.7ex\hbox{$
\;\stackrel{\textstyle<}{\sim}\;$}}

\begin{document}

\begin{titlepage}

\begin{flushright}
FTPI-MINN-11/28, UMN-TH-3020/11\\
\today
\end{flushright}

\vspace{1cm}

\begin{center}
{  \Large \bf  \boldmath{$\mathcal N=(0,2)$}  Deformation of \boldmath{$\mathbf{CP}(1)$} Model: Two-dimensional Analog of \boldmath{$\mathcal{N}=1$} Yang-Mills Theory in Four Dimensions }
\end{center}

\vspace{0.3cm}

\begin{center}
{\large
Xiaoyi Cui$^{\,a}$ and M.~Shifman$^{\,a,b}$}
\end {center}

\vspace{1mm}

\begin{center}

$^{a}${\it  Department of Physics, University of Minnesota,
Minneapolis, MN 55455, USA}\\[1mm]
$^b${\it  William I. Fine Theoretical Physics Institute,
University of Minnesota,
Minneapolis, MN 55455, USA}

\end{center}

\vspace{0.2cm}

\begin{center}
{\large\bf Abstract}
\end{center}

\vspace{0.2cm}

We consider two-dimensional $\mathcal{N}=(0,2)$ sigma models with the $\mathbf{CP}(1)$ target space. A minimal model of this type 
has one left-handed fermion. Nonminimal extensions contain, in addition,  $N_f$ right-handed fermions. 
Our task is to derive expressions for the $\beta$ functions valid to all orders. To this end we use a variety of methods:
(i) perturbative analysis; (ii) instanton calculus; (iii) analysis of the supercurrent supermultiplet (the so-called hypercurrent) and 
its anomalies, and some other arguments. All these arguments, combined, indicate a direct parallel between 
the heterotic  $\mathcal{N}=(0,2)$ $\mathbf{CP}(1)$ models and four-dimensional super-Yang--Mills theories. In particular, the minimal 
$\mathcal{N}=(0,2)$ $\mathbf{CP}(1)$ model is similar to ${\mathcal N}=1$ supersymmetric gluodynamics. Its exact $\beta$
function can be found; it has the structure of the Novikov--Shifman--Vainshtein--Zakharov (NSVZ) $\beta$ function of supersymmetric gluodynamics. The passage to nonminimal  $\mathcal{N}=(0,2)$ sigma models is equivalent to adding matter. In this case an
NSVZ-type exact relation between the $\beta$ function and the anomalous dimensions $\gamma$ of the ``matter" fields
is established. We derive an analog of the Konishi anomaly. At large $N_f$ our $\beta$ function develops an infrared
fixed point at small values of the coupling constant (analogous to the Banks--Zaks fixed point). Thus, we reliably predict the existence of a conformal window. At $N_f=1$ the model under consideration reduces to the well-known $\mathcal{N}=(2,2)$ $\mathbf{CP}(1)$ model.

\end{titlepage}

\newpage

\tableofcontents

\newpage

\section{Introduction}

This paper could have been called ``Perturbative and nonperturbative aspects of $\mathcal{N}=(0,2)$ sigma models: the $\beta$ 
function, Konishi anomaly, conformal window and all that in $\mathbf{CP}(1)$." 2D-4D correspondence is a popular topic in
the current literature. Its discussion has 
a long history, see e.g.  \cite{HT1,ND98,ABEKY,SYmon,HT2, Dorey:2011pa, Cecotti:2010fi}. Most theoretical efforts were focused
 on a relation between four-dimensional $\mathcal{N}=2$ SQCD and  two-dimensional $\mathcal{N}=(2,2)$ sigma models. 
 The former support non-Abelian strings \cite{HT1,ABEKY}. The latter appear as
 low-energy effective theories on the non-Abelian string world sheet. It is not surprising then that the BPS-protected sectors
 of the 4D parents and 2D daughter theories are related.
  For more details on this, the readers are referred to \cite{Trev}.
  
 Later on, the bulk theories supporting non-Abelian strings were deformed to break ${\mathcal N}=2$ in 4D down to 
 ${\mathcal N}=1$. It was found \cite{EdTo,SY1} that the low-energy theories on the string world sheet are no longer 
 $\mathcal{N}=(2,2)$ supersymmetric. Instead, one gets $\mathcal{N}=(0,2)$ heterotic sigma models with the
 $\mathbf{CP}(N-1)$ target space. This finding gave a strong impetus to explorations of these heterotic models which had been 
 previously discussed only in general terms \cite{Adams:2003zy, Witten05, hep-th/0604179, TY1, Frenkel:vz}.
  
In this paper we will study two-dimensional $\mathcal{N}=(0,2)$ sigma models with the $\mathbf{CP}(1)$ target space. 
A minimal model of this type 
has one left-handed fermion which, together with a complex scalar field,
enters an $\mathcal{N}=(0,2)$ chiral superfield. This minimal model can be readily extended.
Nonminimal extensions contain, in addition,  $N_f$ right-handed fermions. 
In particular, if $N_f=1$, the nonminimal model under consideration reduces to the conventional
 $\mathcal{N}=(2,2)$ $\mathbf{CP}(1)$ model.

In this paper  we will focus on various   derivations
of exact expressions  for the $\beta$ functions (valid to all orders in the $\mathbf{CP}(1)$ coupling). 
Remarkably, our results will exhibit a direct parallel between 
the heterotic  $\mathcal{N}=(0,2)$ $\mathbf{CP}(1)$ models and four-dimensional super-Yang--Mills theories. In particular, the minimal 
$\mathcal{N}=(0,2)$ $\mathbf{CP}(1)$ model is similar to ${\mathcal N}=1$ supersymmetric gluodynamics. Its exact $\beta$
function can be found; it has the structure of the Novikov--Shifman--Vainshtein--Zakharov (NSVZ) $\beta$ function 
\cite{NSVZ2,NSVZ3} in 
  supersymmetric Yang--Mills theory without matter. Then we pass  to nonminimal  $\mathcal{N}=(0,2)$ sigma models.
  It turns out that this passage corresponds to adding (adjoint) matter in 
  four-dimensional super-Yang--Mills theory. Thus, in the nonminimal $\mathcal{N}=(0,2)$ $\mathbf{CP}(1)$ models
we will obtain an 
NSVZ-type exact relation between the $\beta$ function and the anomalous dimensions $\gamma$ of the ``matter" fields.

Our arguments will be based on a number of methods.
First, we will carry out a perturbative (super)graph analysis. This will allow us to obtain the
$\beta$ functions at the two-loop level. Comparison with the $N_f=1$ case which is in fact $\mathcal{N}=(2,2)$
will give us the first indication on the emergence of the NSVZ-type
$\beta$ function.

Then we will study the instanton measure, using parallels 
with the analogous NSVZ derivation.
We will obtain a version of  the nonrenormalization theorem in the instanton background. Essentially we will
demonstrate that the instanton measure is exhausted by a one-loop calculation,
in much the same way as was the case in  4D super-Yang--Mills theories \cite{NSVZ2} and in 2D $\mathcal{N}=(2,2)$   sigma models
\cite{Morozov:1984ad}. From this result one can readily deduce a $\beta$ function of the
NSVZ type.

Our third argument is based on the analysis of the supercurrent supermultiplet (the so-called hypercurrent) and 
its anomalies. Not only will the NSVZ $\beta$ function be confirmed, but, in addition we will 
understand the difference between the holomorphic and canonic couplings, which is exactly the same as in
the 4D super-Yang--Mills \cite{SV91}. {\em En route} we will derive a 2D  analog of the
Konishi anomaly. This is a necessary element of the $\beta$ function derivation through the hypercurrent anomaly.
The exact formula that we obtain relates the $\beta$ function of the nonminimal models with the anomalous dimension of the
``matter fields." The latter is known as an expansion in perturbation theory. 

At large $N_f$ our $\beta$ function develops an infrared
fixed point at small values of the coupling constant (analogous to the Banks--Zaks fixed point
\cite{BZ}\,\footnote{More exactly, it should have been referred to as the Belavin--Migdal--Banks--Zaks.}). 
Since the position of this fixed point is at $g^2 \sim 1/N_f$, we can use the leading-order result
for the anomalous dimension to {\em prove}
the existence of the fixed point. In other words, in the nonminimal models a conformal window exists starting from some critical value $N_f^*$. Near the lower edge of the conformal window the theory is presumably strongly
coupled.

One can ask a natural question: Why do we consider only the $\mathbf{CP}(1)$ model and do not generalize to CP($N-1)$
with arbitrary $N$? This is due to
an anomaly in heterotic models pointed out in \cite{MN}. 
This anomaly prevents us from considering the models we study in this paper for arbitrary $N$.
However, some other nonminimal generalization of the $\mathcal{N}=(0,2)$ CP($N-1)$ models
will be studied  in our forthcoming work \cite{CS4}.

The structure of this paper is as follows. We formulate 
the minimal $\mathcal{N}=(0,2)$ models in Sec.~\ref{sec:1}. In Sec.~\ref{sec:superf} we 
carry out  perturbative calculations of the $\beta$ function
  up to two-loop order, in superfield formalism, as outlined in \cite{CS2}. In Sec.~\ref{sec:inst} we start studying nonperturbative effects in the minimal model (instanton and its measure).  We construct exact instanton measure. In this construction we take into account zero modes, one-loop effects in the instanton background,
  and then,  following NSVZ \cite{NSVZ},   use a nonrenormalization theorem for two and more loops.
  The instanton background gives us a particularly clear way to see the cancellation of higher loops.
The all-loop exact $\beta$ function is presented in Sec.~\ref{sec:nsvz}.  In Sec.~\ref{sec:current} we calculate explicitly the supercurrent supermultiplet for this model. In Sec.~\ref{sec:2} we extend the minimal model by adding ``matter", i.e. the right-handed fermion fields. Following the same road as in the minimal model,
we calculate the two-loop $\beta$ function perturbatively in the nonminimal model.
Then we exploit the instanton analysis to obtain an exact relation between the $\beta$ function and the anomalous dimension $\gamma$ of the ``matter" fields. In Sec.~\ref{hcnf} we calculate the supercurrent supermultiplets for the extended (nonminimal) models. Section~\ref{konishi} is devoted to a 2D analog of the Konishi anomaly in the extended models. Finally, Sec.~\ref{conwid} demonstrates  the appearance of a conformal window. 
Main conclusions and prospects for future explorations are summarized in Sec.~\ref{conclu}.

\section{Formulation of the minimal heterotic $\mathbf{CP}(1)$ \\
model}
\label{sec:1}

In this section we will formulate the minimal  $\mathcal{N}=(0,2)$ $\mathbf{CP}(1)$ sigma model (previously it was studied e.g. 
in \cite{Witten05, TY1}).  We will use $\mathcal{N}=(0,2)$ superfield formalism.
Note that due to the anomaly in \cite{MN} it is impossible to generalize this model to $\mathbf{CP}(N-1)$.

The Lagrangian of the model under consideration is 
\beq
\mathcal{L}_A = \frac 1{g^2} \int d^2 \theta_R
\frac{A^\dagger i\overset{\leftrightarrow}{\partial}_{RR} A}{1+A^\dagger A}\,, 
\eeq
where $A$ is a bosonic chiral superfield:
\beq
A(x,\theta_R^\dagger, \theta_R) = \phi(x)+\sqrt{2}\theta_R\psi_L(x)+i\theta_R^\dagger\theta_R\partial_{LL}\phi\,,
\eeq  
$\phi$ is a complex scalar, and $\psi_L$ is a left-handed Weyl fermion. 
We define $\overset{\leftrightarrow}{\partial}_{RR}$ to be $\overset{\rightarrow}{\partial}_{RR}/2 - \overset{\leftarrow}{\partial}_{RR}/2$.
The superfield $A$ can be understood as taking values on the $\mathbf{CP}(1)$ manifold, and, thus, can be endowed with the following nonlinear transformations:
\beq
A\to A+\epsilon +\bar \epsilon A^2\,,\quad A^\dagger\to A^\dagger +\bar \epsilon +\epsilon (A^\dagger)^2\,,
\eeq
plus a U$(1)$ rotation. 

In components, we can write the Lagrangian as
\beq
G\left\{
\partial^\mu\phi\partial_\mu\phi^\dagger+i\psi_L^\dagger \overset{\leftrightarrow}{\partial}_{RR} \psi_L -
{2i}  \,\frac{1}{\chi}\, \psi_L^\dagger \psi_L\,\phi^\dagger\overset{\leftrightarrow}{\partial}_{RR}\phi
\right\}\,.
\label{4}
\eeq
The derivatives $\partial_{RR}$ and $\partial_{LL}$ are defined in Appendix A, see Eq. (\ref{72}).
Here we denote by $G$
the K\"ahler  metric on the target space ($S^2$ in the case at hand), in the Fubini--Study form,
\beq
G =
\frac{2}{g^2\,\chi^{2}}\,,
\label{kalme}
\eeq
where
\beq
\chi \equiv 1+\phi\,\phi^\dagger\,.
\label{chidef}
\eeq
Moreover,
$R$ is the Ricci tensor,
\beq
 R =\frac{2}{\chi^2}\,,
\label{Atwo}
\eeq
while $g^2$ is the coupling constant. 

The coupling constant $g$ can be complexified. In what follows we will deal with the holomorphic coupling
$g_h$ defined as
\beq
\frac{2}{g_h^2} = \frac 2{g^2}+i\frac{\omega}{2\pi}\,.
\label{holomorcoup}
\eeq
In terms of the holomorphic coupling the Lagrangian of the
minimal model has the form
\beqn
\mathcal{L}_A
&=&
\int d^2\theta_R \frac i{2g_h^2}\frac {A^\dagger \partial_{RR} A}{1+A^\dagger A} + {\rm H.c.}
\nonumber\\[3mm]
&=&
-\frac i{2g_h^2}\int d\theta_R \frac {\bar D_L A^\dagger \partial_{RR} A}{(1+A^\dagger A)^2} + \frac i{2\bar{g}_h^2} \int d\theta_R^\dagger \frac {D_L A \partial_{RR} A^\dagger}{(1+A^\dagger A)^2}\,.
\label{h}
\eeqn
The target space  invariance of the integrand is maintained in the second line.
In perturbative loop calculations and in  instanton analysis
we will use the canonical coupling $g$. To differentiate between the bare and renormalized couplings
we will use subscripts 0 and r where appropriate.

In Sect. \ref{sec:2}
we will extend this minimal model by adding $N_f$ ``matter" fields.

\section{Perturbative    superfield calculation \\of the 
\boldmath{$\beta$} function   }
\label{sec:superf}

Fermions do not contribute to the $\beta$ function at one loop (see e.g. \cite{NSVZ}). Therefore, the first coefficient of the
$\beta$ function in the minimal heterotic model is the same as in the nonsupersymmetric $\mathbf{CP}(1)$ model (see \cite{NSVZ, CS1}).
The first nontrivial task to address is the calculation of the second coefficient.

In this section we will use the
superfield method to calculate the two-loop $\beta$ function in the minimal model. 
We will use a linear background field method, setting the background field $$A_{bk}=fe^{-i x\cdot k}\,.$$ The basic method is roughly the same as that in \cite{NSVZ}. The superfield calculation was outlined
 in our previous paper \cite{CS2}. We expand the action around the 
 chosen background, splitting the superfield $A$ into two parts, classical (background) and quantum.
Then we  calculate relevant diagrams with quantum fields in loops. 

\begin{figure}
\begin{center}
\includegraphics[width=5in]{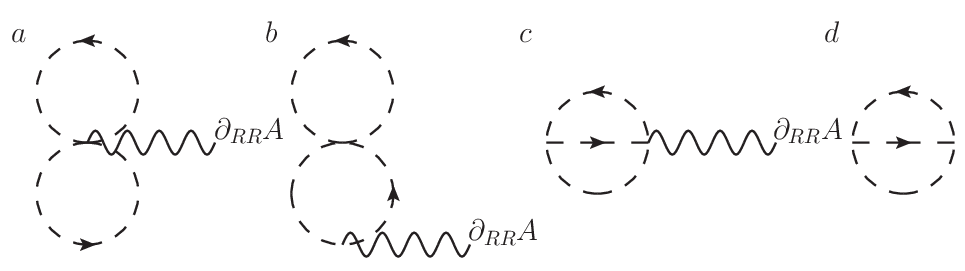}
\end{center}
\caption{\footnotesize Two-loop correction to the coupling $g$ by $A$-loops. The dashed lines denote the propagator of the quantum part of $A$ in the chosen background, while the wavy lines denote the background field.}
\label{super2loop_111024}
\end{figure}

If we limit ourselves to the origin in the target space (i.e. $\phi =0$) and forgo the check of the target space invariance , 
at two-loop order the $\beta$ function is determined by the diagrams in Fig.~\ref{super2loop_111024}.
As previously we use   the $\epsilon$ regularization, where $$\epsilon = 2-d\,.$$

The last diagram does not explicitly exhibit $\partial_{RR} A$ as the external line, but it does produce a 
 contribution due to momentum insertion. A quick evaluation tells us that the first three diagrams contribute only 
 double poles, and they  cancel  among themselves, as they should. The graph-by-graph results are listed in Table~\ref{TB_Aloop},
 where the following notation is used
 \beq
 I\equiv\int \frac{d^{2-\epsilon}p}{(2\pi)^d}\,\,\frac1{p^2-m^2}\,.
 \label{8}
 \eeq
 In logarithmically divergent graphs the following correspondence takes place at one loop:
 \beq
 \frac{1}{\epsilon}\, \leftrightarrow \,\text{ln}\frac{M}{m}\,,
 \label{mon9}
 \eeq
where the left-hand side represents dimensional regularization, while the right-hand side the Pauli--Villars regularization; $M$ is the mass of the Pauli--Villars regulator.
\begin{table}
\begin{center}
\begin{tabular}{| c || c | c |}
\hline
Diagram & Double pole & Single pole \rule{0mm}{6mm}\\[2mm] \hline
a &  $-\frac{3}{2}\frac{g^2}2 \frac{A^\dagger i\partial_{RR} A}{1+A^\dagger A} I^2+$H.c. & $0$\rule{0mm}{6mm}\\[2mm] 
b &  $\frac{g^2}{2}  \frac{A^\dagger i\partial_{RR} A}{1+A^\dagger A} I^2+$H.c. & $0$ \rule{0mm}{6mm}\\[2mm] 
c & $\frac 12\frac{g^2}2  \frac{A^\dagger i\partial_{RR} A}{1+A^\dagger A} I^2+$H.c. & $0$\rule{0mm}{6mm}\\[2mm]
d & $0$ & $-\frac{g^2}2\frac{i}{4\pi}  \frac{A^\dagger i\partial_{RR} A}{1+A^\dagger A} I+$H.c. \rule{0mm}{6mm}\\[2mm] 
\hline
\end{tabular}
\end{center}
\caption{\footnotesize Two-loop calculation of $g^{-2}$ renormalization in the $\epsilon$ expansion follows that in Figure \ref{super2loop_111024}. $I=\int \frac{d^{2-\epsilon}p}{(2\pi)^d}\frac1{p^2-m^2}$, see \cite{CS1}.}
\label{TB_Aloop}
\end{table}
The remaining contribution due to the last diagram results in the following two-loop $\beta$ function:
\beq
\frac 1{g_r^2} = \frac 1{g_0^2}\left(
1-ig_0^2I-\frac i{4\pi} g_0^4 I
\right),
\eeq
or 
\beq
\beta(g^2) = -\frac{g^4}{2\pi}\left( 1+\frac{g^2}{4\pi} \right)\,.
\label{betacp102}
\eeq

Below we will argue that higher loops iterate the two-loop expression in a geometrical progression, so that
the full result for the $\beta$ function in the minimal heterotic model is
\beq
\beta(g^2) = -\frac{g^4}{2\pi}\left( 1-\frac{g^2}{4\pi} \right)^{-1}\,.
\eeq
\label{betacp102p}
A parallel with the NSVZ $\beta$ function in supersymmetric gluodynamics \cite{NSVZ2,NSVZ3} is evident.

\section{Non-perturbative calculation through the instanton measure}
\label{sec:inst}

Bosonic $\mathbf{CP}(N-1)$ models exhibit instanton solutions \cite{Pol,Per}. Hence, this is also the case for 
the $\mathcal{N}=(0,2)$ models. For $\mathbf{CP}(1)$, the bosonic (anti-)instanton solution with the unit topological charge is
\beq
\phi = \frac{y}{z-z_0}\,,\quad \phi^\dagger = \frac{\bar y}{\bar z-\bar z_0}\,,
\eeq 
where $y$ and $z_0$ are the collective coordinates: $z_0$ is the instanton center
while a complex number $y$ parametrizes its size and
a U(1) phase.  Our notation in Euclidean space-time is explained  in Appendix B to which  the reader
is referred for further details. We easily get the bosonic zero modes, by taking derivatives of the instanton solution with respect to the above collective coordinates. There are four (real) zero modes, or, two complex \cite{NSVZ}.

The fermion zero modes can be obtained by applying supersymmetry and superconformal symmetry. From the supersymmetry transformation induced by $Q^\dagger$, one obtains the following fermion zero mode:
\beq
\psi_{\bar z} ^\dagger = \frac{\bar y\alpha}{(\bar z-\bar z_0)^2}\,.
\eeq

From the superconformal transformation, we get another zero mode,
\beq
\psi_{\bar z} ^\dagger = \frac{\bar y\beta^\dagger}{\bar z-\bar z_0}\,.
\eeq
Note that in the $\mathcal{N}=(0,2)$ theory we deal with {\em two}
 fermion zero modes rather than four, which appear in the $\mathcal{N}=(2,2)$ $\mathbf{CP}(1)$ model. The reason is that, involution is lost upon transition to Euclidean space. No zero mode arises from the background 
 $\phi = \frac{y}{z-z_0}$ (see also \cite{TY1}). 
 This means that the superinstanton under consideration has no collective coordinates
 $\alpha^\dagger$ and $\beta$.
 The fact that we deal with two rather than four fermion zero modes agrees with the coefficient in the chiral anomaly (see Sec.~\ref{sec:current}) which is twice smaller in $\mathcal{N}=(0,2)$ compared to 
 $\mathcal{N}=(2,2)$.

Assembling everything together, we obtain the instanton superfield in the form
\beq
A_{\rm inst} = \frac{y}{z-z_0}\,,\quad A^\dagger_{\rm inst} = \frac{\bar y(1+4i\theta^\dagger \beta^\dagger)}{\bar z_{\rm ch}-\bar z_0-4i\theta^\dagger\alpha}\,,
\eeq
where\,\footnote{Note that in Sect. \ref{sec:inst} we will use $\theta$ and $\theta^\dagger$ to denote the Grassmannian variables in Euclidean superspace. We intentionally drop the subscript ``$R$" to distinguish from those in Minkowski superspace. } $$\bar z_{\rm ch} = \bar z-2i\theta^\dagger\theta\,.$$

To derive the instanton measure, we need to define the integral over the collective coordinates. 
To this end, as usual, we proceed from the mode expansion to the collective coordinates of the zero modes
(moduli). In particular, as explained in \cite{NSVZ}, we need to calculate the normalization of the zero modes given by 
\beq
\int dz d\bar z\,\,  G_{ij}\delta\phi_i\delta\phi_i^\dagger\,.
\label{dev}
\eeq
As a technical point, we note that two of the bosonic (real) modes (conformal) and the fermionic superconformal mode are actually logarithmically divergent in the infrared under the normalization. However, these divergences are canceled by 
similar divergences coming from the one-loop contribution due to the nonzero modes. This was  explicitly
verified  in the  case of nonsupersymmetric $\mathbf{CP}(1)$ models in \cite{Fateev:1979dc};  the argument 
readily extends to the supersymmetric case too. 

As it follows from the norm of the modes, each (complex) boson zero mode is accompanied by the factor 
$ 2/{g^2}$ and each (complex) fermion zero mode is accompanied by the factor $ {g^2}/2$. The dependence on the instanton size $|y|$ will be omitted temporarily and recovered at a later stage on the basis of
 dimension arguments. Hereafter, we will drop the constant numerical factors, since they contribute only to an overall constant.
As a result, at this stage we arrive at the following instanton measure
\beq
d\mu=\text{ const. } \left(\frac1{g^2}\right)^{n_b}\left(g^2\right)^{n_f}\,e^{-\frac{4\pi}{g^2}}\,\, dy d\bar y \,\, \,dz_0 d\bar z_0 \, d \alpha d\beta^\dagger\,,
\label{dvad}
\eeq
where $n_b=2$ and $n_f=1$. (We hasten to add that this is {\em not} the final result.)

So far quantum corrections have not yet been discussed. In the $\mathcal{N}=(2,2)$ model, the one-loop corrections due to the nonzero modes  in the instanton background cancel each other  completely \cite{NSVZ, Morozov:1984ad}. In the $\mathcal{N}=(0,2)$ model this is not quite the case.  Let us consider the one-loop effects in more detail. For the nonzero bosonic modes, we will expand the field $\phi$ as
\beq
\phi = \phi_{\rm inst}+\frac g{\sqrt{2}}\delta \phi=\phi_{\rm inst}+\frac g{\sqrt{2}}\sum_n \phi_na_n\,.
\label{expa}
\eeq
Note that the part $\phi_{\rm inst}$ contains   the boson zero modes.
The functions $\phi_n$ in the expansion (\ref{expa}) are the eigenfunctions of the operator 
\beq
-\frac\partial{\partial z} \frac 1{\chi_{\rm inst}^2}\frac\partial{\partial \bar z}\phi_n = E_n^2\,\, \frac{\phi_n}{\chi_{inst}^2}
\eeq
normalized by the condition
\beq
\int \frac{\phi_n^\dagger \phi_n}{\chi_{inst}^2} d^2x = 1\,,
\eeq
where $\chi_{\rm inst} = 1+\phi_{\rm inst}^\dagger\phi_{\rm inst}$, and $E_n^2$ is the $n$-th eigenvalue.
At the one-loop level we can rewrite  the action as 
\beq
-\frac{4\pi}{g^2}-\int d^2 x \sum_n E_n^2a_n^\dagger a_n\frac {\phi_n^\dagger\phi_n}{\chi_{\rm inst}^2}\,,
\eeq
which, according to the standard rule of functional integration, gives
\beq
\int [\mathscr{D}\delta\phi][\mathscr{D}\delta\phi^\dagger]\rightarrow \left\{\det \left[ -\frac\partial{\partial z} \frac 1{\chi_{inst}^2}\frac\partial{\partial \bar z} \right]\right\}^{-1} = \prod_n\frac{1}{E_n^2}\,.
\eeq

As for the fermion nonzero modes, we perform  a similar expansion. Note that after the Wick rotation, the
left-handed fermion $\psi_L$ is no longer related to $\psi_L^\dagger$ by Hermitian conjugation. 
Therefore in this section we will use $\psi_{\bar z}$ and $\psi_z^\dagger$,  respectively, to denote them. 

Consider the expansion for the fermion fields
\beq
\psi_{\bar z} = \psi_{\bar z,{\rm inst}}+\frac{g}{\sqrt{2}} \sum_n b_n u_n\,,\quad 
\psi_z^\dagger = \psi_{z,{\rm inst}}^\dagger+\frac{g}{\sqrt{2}} \sum_n c_n \bar v_n\,,
\eeq
where $b_n$ and $c_n$ are complex Grassmannian parameters. The functions $u_n$ and $v_n$ are defined via
\beq
i\partial_z\frac 1{\chi_{\rm inst}^2} u_n = \frac{\mathcal{E}_n}{\chi_{\rm inst}^2} v_n\,,\quad 
i\partial_{\bar z} v_n  = \mathcal{E}_n u_n\,,
\label{fnzmodeseqn}
\eeq
subject to the normalization conditions similar to that of $\phi_n$.
The part of the action that contains $\psi_L$ now becomes 
\beq
\int d^2 x \, \,i \psi_z^\dagger \partial_{z} \frac{2}{g^2\chi_{\rm inst}^2}\psi_{\bar z}=\int d^2 x\,\,  i
\sum_n \mathcal{E}_n c_n b_n\frac{\bar v_n v_n}{\chi_{\rm inst}^2}\,.
\eeq
Therefore, integration over the Grassmannian parameters yields $\prod_n \mathcal{E}_n$.
 Note that in   solving  Eq.~(\ref{fnzmodeseqn}) we obtain 
\beq
-\partial_z\frac{1}{\chi_{\rm inst}^2} \partial_{\bar z}v_n = \frac{\mathcal{E}_n^2}{\chi_{\rm inst}^2} v_n\,,
\eeq
which is exactly the same as the equation that defines $\phi_n$. Hence the boson-fermion degeneracy follows, $$\mathcal{E}_n^2 = E_n^2\,.$$ 

In principle the eigenvalue $\mathcal{E}_n$ could be both positive and negative. 
Let us elucidate this subtle point.
In fact, here we are double counting the eigenstates in calculating the fermion determinant. It is easy to 
see that this is the case if we turn first to the  $\mathcal{N}=(2,2)$ theory. There, the relevant action is given by
\beq
\int d^2 x\,\left(  i
\sum_n \mathcal{E}_n c_n b_n\frac{\bar v_n v_n}{\chi_{\rm inst}^2}+ i 
\sum_n \mathcal{E}_n \bar b_n \bar c_n\frac{\bar u_n u_n}{\chi_{\rm inst}^2}\right).
\eeq
Integration runs over four Grassmann parameters (at each level),
\beq
\prod_n db_n d\bar b_n dc_n d\bar c_n\,,
\eeq
In the $\mathcal{N}=(0,2)$ case  we have to identify
\beq
b_n \leftrightarrow  \bar c_n\,,\quad  c_n \leftrightarrow \bar b_n\,.
\eeq
In other words,  we should count only the field configurations that correspond either to $\mathcal{E}_n$  or to  $-\mathcal{E}_n$ (assuming $E_n>0$). 
 
As a result, in our $\mathcal{N}=(0,2)$ theory, $\prod \mathcal{E}_n$ should be understood as 
$(\prod E_n^2)^{\frac 12}$, and, hence, symbolically we can write
\beq
\int [\mathscr{D}\delta\psi_{\bar z}][\mathscr{D}\delta\psi_z^\dagger]\rightarrow \left\{\det \left[\begin{array}{cc}0 & i\partial_z\frac{1}{\chi_{\rm inst}^2} \\[1mm]
i\frac{1}{\chi_{\rm inst}^2}\partial_{\bar z} & 0\end{array}\right] \right\}^{\frac12}=\left(\prod_n E_n^2\right)^{\frac12}\,.
\label{tritri}
\eeq
The product runs over the nonzero modes.

As a result, due to the lack of balance between the numbers of the modes (bosonic versus fermionic), we 
do not have 
complete cancellation of the one-loop correction coming from the boson nonzero modes by that coming from the fermion nonzero modes. 
This is in contradistinction with the situation in the $\mathcal{N}=(2,2)$ theory.

We  have to evaluate the one-loop contribution from the
nonzero modes in the instanton background.  In four dimensions this kind of calculation was carried out  in \cite{tHooft}, and in the pure bosonic $\mathbf{CP}(1)$ model it has been done in \cite{Fateev:1979dc}. All we need to know is the general form of 
the one-loop correction due to nonzero modes in the instanton measure, 
$\exp\left( {\rm const. }\log\frac{M}{|y|}\right)$, with no explicit $g^2$ dependence.\footnote{See, however, a remark after Eq. (\ref{dev}).}
Here $M$ is mass of the ultraviolet (UV) regulator of the theory.  Thus, the one-loop effect will bring us a prefactor $M^\kappa$. 
We will determine it using our knowledge of the bosonic $\mathbf{CP}(1)$ model. 

Explicitly, we can write down the instanton measure for the bosonic 
 $\mathbf{CP}(1)$ model to one-loop order, which is given in \cite{Fateev:1979dc} and also entirely fixed by the $\beta$ function at the two-loop level. We will postpone the second derivation till Sec.~\ref{sec:nsvz}, and just show the final result. The measure is 
\beq
d\mu \sim \left(\frac{M^2}{g^2}\right)^{n_b}M^{-2}\,\, dy d\bar y\,\,  dz_0 d\bar z_0\,,\qquad n_b =2\,,
\eeq
where the factor $M^{-2}$ comes from the one-loop correction due to the nonzero modes, and, hence, $\prod_n E_n^{-2} = M^{-2}$. Given Eq.~(\ref{tritri}), we immediately conclude
that  the one-loop correction to the instanton measure  in our $\mathcal{N}=(0,2)$ model is $M^{-1}$. 

With this knowledge in hand we can return to
Eq.~(\ref{dvad}) which contains only  zero modes. After inserting nonzero mode one-loop effects
we find the instanton measure in the form
\beq
d\mu\sim \left(\frac{M^2}{g^2}\right)^{n_b}\left(\frac{g^2}{M}\right)^{n_f}M^{-1}e^{-\frac{4\pi}{g^2}}\,\,
 dy d\bar y\,\,  dz_0 d\bar z_0 d \,\, \alpha d\beta^\dagger\,,
\eeq
with $n_b=2$ and $n_f=1$. As we will argue in Sect. \ref{sec:nsvz}, this is the exact formula.

Finally, note that the instanton measure is dimensionless. Therefore, 
we need to reinstate an appropriate  dimensional parameter. There is a unique   choice, the instanton size, which is,
simultaneously,  the infrared cutoff in the instanton calculation. It is given by $|y|$. 

This leaves us with the following master formula for the measure in the $\mathcal{N}=(0,2)$ $\mathbf{CP}(1)$ model:
\beqn
d\mu
&=& 
\left(\frac{M^2 }{g^2}\right)^{n_b}\left(\frac{g^2}{M}\right)^{n_f}(M)^{-1}\, e^{-\frac{4\pi}{g^2}}
\,\,  d\text{log}(y) d\text{log}(\bar y)\,\,  dz_0 d\bar z_0 \,\, d \alpha d\beta^\dagger\,,
\nonumber\\[3mm]
n_b &=& 2\,,\quad n_f=1\,.
\label{imf}
\eeqn

\subsection{A nonrenormalization theorem}
\label{sec:nsvz}

This is not the end of the story, however. We have to address the question of two- and higher-loop corrections in the instanton background. In this subsection we will argue that they vanish. Our arguments are 
intended to  show that the instanton measure   in (\ref{imf}) is all loop exact, i.e., it does not receive higher-loop corrections. The proof is a version of the nonrenormalization theorem \cite{NSVZ2, CS2}. 

Let us recall that in the instanton background superfield $A_{\rm inst}$ and $A^\dagger_{\rm inst}$, we can apply supersymmetry transformation  given by 
\beq
\theta \to \theta +\epsilon\,,\quad \theta^\dagger \to \theta^\dagger+\epsilon^\dagger\,,\quad \bar z_{\rm ch}\to \bar z_{\rm ch}+4i\epsilon\theta^\dagger\,. 
\label{est}
\eeq
Under such a transformation the superinstanton transforms as
\beqn
A_{\rm inst} &=& \frac y{z-z_0}\to \frac y{z-z_0}\,,\\[2mm]
A_{\rm inst}^\dagger &=& \frac{\bar y(1+4i\theta^\dagger \beta^\dagger)}{\bar z_{\rm ch}-\bar z_0-4i\theta^\dagger\alpha} \to  \frac{\bar y(1+4i\theta^\dagger \beta^\dagger+4i\epsilon^\dagger\beta^\dagger)}{\bar z_{\rm ch}-\bar z_0+4i \epsilon\theta^\dagger-4i\theta^\dagger\alpha-4i\epsilon^\dagger\alpha}\,.
\eeqn
To make the background invariant under such transformations, one can assign appropriate  transformation laws to the collective coordinates (moduli), namely,
\beq
\bar y\to \bar y(1+4i\epsilon^\dagger\beta^\dagger)\,,\quad \bar z_0\to \bar z_0+4i\epsilon^\dagger \alpha\,,\quad\alpha\to \alpha+\epsilon\,,\quad \beta^\dagger\to \beta^\dagger\,.
\label{collecoortra}
\eeq
Combining  (\ref{est}) and (\ref{collecoortra}) it is not difficult to see that the instanton field configuration remains intact when we apply the supersymmetry transformations. Moreover, our expression for the integration measure over the collective coordinates, (\ref{imf}),  is invariant too. This implies the following. Supersymmetry
understood as a combined action of (\ref{est}) and (\ref{collecoortra}) is preserved classically by our chosen instanton background, and, hence, it will be preserved in loops. In particular, this forbids any correction 
to (\ref{imf}) of the form\,\footnote{The scale $|y|$ serves as a natural infrared cutoff. Note that the infrared cutoff is provided by the absolute value  of $y$, while the dependence on the phase angle is trivial.} $1+c g^2 \text{ log}(M^2|y|^2)$, since such a term would change the power of $\bar y$, that would be in contradiction with (\ref{collecoortra}). 

Moreover, nonlogarithmic corrections of the type $1+cg^2+c'g^4+\cdots$ also do not show up in multiloop calculations. This follows from a nonrenormalization argument similar to \cite{CS2}.
 Consider a correction that could possibly come from two or more loops. In this case we can always write the loop integration in the   form
\beq
\int d^2z d\theta d\theta^\dagger 
f(x, \theta, \theta^\dagger, z_0, \bar z_0, y, \bar y,  \alpha, \beta^\dagger)\,,
\eeq
where the function $f$ must be invariant under supersymmetry transformations supplemented by (\ref{collecoortra}). This tells us that $f$ can only be a function of the following (invariant) arguments:
\beq
y\,,\quad \bar y(1+4i\beta^\dagger\theta^\dagger)\,,\quad z-z_0\,,\quad \bar z_{\rm ch}-\bar z_0+4i\theta^\dagger \alpha\,,\quad \theta-\alpha\,, \quad \beta^\dagger\,.
\eeq
In the subsequent analysis we will only indicate the explicit dependence of $f$ on $\bar y(1+4i\beta^\dagger\theta^\dagger)$, $\bar z_{\rm ch}-\bar z_0+4i\theta^\dagger \alpha$, $\theta-\alpha$ and $\beta^\dagger$. 
Only these variables will be of importance. Due to the Grassmannian nature of $\theta-\alpha$ and $\beta^\dagger$, the function $f$ can be represented as 
a sum of two terms, 
\beqn
&&f\left(\bar y(1+4i\beta^\dagger\theta^\dagger), \bar z_{ch}-\bar z_0+4i\theta^\dagger \alpha, \theta-\alpha, \beta^\dagger\right) \nonumber\\[2mm]
&=& f_0\left(\bar y(1+4i\beta^\dagger\theta^\dagger), \bar z_{ch}-\bar z_0+4i\theta^\dagger \alpha\right) \nonumber\\[2mm]
&&+ (\theta-\alpha)\beta^\dagger f_1\left(\bar y(1+4i\beta^\dagger\theta^\dagger), \bar z_{ch}-\bar z_0+4i\theta^\dagger \alpha\right),
\eeqn
where $f_{0,1}$ are some other functions.
It is obvious that upon integration over $\theta$, only $f_1$ can survive, and the integration takes the form 
\beq
\int d^2 z \,\, d\theta^\dagger \, \beta^\dagger \, f_1\left(\bar y, \bar z-\bar z_0+4i\theta^\dagger \alpha\right).
\eeq

Next, we shift $\bar z$, and then the remaining integral has to vanish. It vanishes, indeed!
 Note that the integration is finite and local, hence the shift in $\bar z$ must be valid.

\subsection{The full \boldmath{$\beta$} function}

Now we know that our expression for the instanton measure is  all-loop exact. 
It depends on the Pauli--Villars regulator mass $M$ explicitly, through $M^2$, and implicitly, through $g^2(M)$. The overall dependence on $M$ must cancel, i.e, 
\beq
\frac{d}{d \text{log}(M)} \left( -\frac{4\pi}{g^2}-\text{log} g^2+\text{log} M^2 \right) = 0\,.
\eeq 
This gives us the all-loop exact $\beta$ function for the coupling constant $g$,
 \beq
 \beta(g^2)=-\frac{g^4}{2\pi}\frac{1}{1-\frac{g^2}{4\pi}}\,.
 \label{39}
 \eeq
 The two-loop coefficient is in agreement with (\ref{betacp102}) determined by a direct perturbation calculation.
 
\section{Supercurrent supermultiplet (hypercurrent)}
\label{sec:current}

In this section we will   analyze the hypercurrent (see \cite{KS,DS}) of the minimal model.
This will set the stage for an
alternative 
derivation of the $\beta$ function which will be completed in Sect. \ref{hcnf}. 
Our consideration will run parallel to that of \cite{NSVZ3}.

Classically, the model under consideration has a conserved U$(1)$ current corresponding
to rotations of the chiral fermion $\psi_L$,  
\beq
j_{LL} =G\psi_L^\dagger\psi_L \,.
\eeq
The supercurrent of ${\mathcal N}= (0,2)$ supersymmetry is  
\beq
S_{LLL} = i\sqrt{2}G\partial_{LL}\phi^\dagger\psi_L\,, \qquad S_{LRR}=0\,.
\eeq
Finally, the energy momentum tensor of our model has the form
\beqn
T_{LLLL} &=& -2G\partial_{LL}\phi^\dagger\partial_{LL}\phi-iG\psi_L^\dagger\mathcal{D}_{LL}\psi_L+iG 
(\mathcal{D}_{LL}\psi_L^\dagger)\psi_L\,,\nonumber\\[2mm]
T_{RRRR} &=& -2G\partial_{RR}\phi^\dagger\partial_{RR}\phi\,,\nonumber\\[2mm]
T_{LLRR}&=&0\,.
\label{51}
\eeqn
It is easy to see that the three currents, $j_{LL}$, $S_{LLL}$ and $T_{LLLL}$ form an $\mathcal{N}=(0,2)$ (non-chiral) supermultiplet, which we will denote by $\mathcal{J}_{LL}$ and refer to it as the hypercurrent,
\beq
\mathcal{J}_{LL} =j_{LL}(x)+i\theta_RS_{LLL}(x)
+i\theta_R^\dagger S_{LLL}^\dagger(x)-\theta_R\theta_R^\dagger T_{LLLL}(x)\,.
\eeq
In fact, the above multiplet has a concise superfield expression, namely, 
\beq
\mathcal{J}_{LL}=\frac12 G(\bar D_L A^\dagger) D_L A\,,
\eeq
with the left-handed fermion current as its lowest component.

As was mentioned, the Lagrangian (\ref{4}) is invariant under U(1) chiral rotations. Therefore,
the current $ {j}_{LL}$ is conserved classically, $\partial_{RR}  {j}_{LL}=0$. 
This also tells us that $j_{RR}=0$. Both relations are certainly true at the classical level.
In fact, it is obvious that at the classical level the hypercurrent 
${\mathcal J}_{LL}$ is conserved as a whole, $\partial_{RR}{\mathcal J}_{LL} = 0$.

 The supercurrent conservation is 
\beq
\partial_{LL}S_{RRL}+\partial_{RR}S_{LLL}=0\,.
\eeq
Classically we have $S_{RRL}=0$, and, therefore, the conservation law simplifies, $\partial_{RR}S_{LLL}=0$. As for the energy-momentum tensor, its conservation tells us that 
\beqn
&\partial_{LL}T_{RRRR}+\partial_{RR}T_{LLRR}=0\,,\nonumber\\[2mm]
&\partial_{LL}T_{RRLL}+\partial_{RR}T_{LLLL}=0\,.
\eeqn
The condition of tracelessness is 
\beq
T_{LLRR}+T_{RRLL}=0\,.
\eeq
Moreover, imposing the ``symmetrycity" condition on $T$,
\beq
T_{LLRR}=T_{RRLL}\,,
\eeq
we obtain that $T_{LLRR}=0$.

\begin{figure}
\begin{center}
\includegraphics[width=2in]{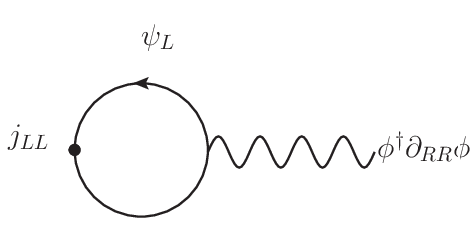}
\end{center}
\caption{\footnotesize One-loop diagram for the $j_{LL}$ anomaly. }
\label{j2ano_111031a}
\end{figure}

Quantum mechanically (i.e. with loops included) the current $j_{LL}$ is anomalous. It is easy to see that the diagram in Fig.~\ref{j2ano_111031a} does not vanish, 
\beq
\partial _{RR} j_{LL}= \frac{i}{2\pi}\frac{\partial_{LL}\phi^\dagger\partial_{RR} \phi-\partial_{RR}\phi^\dagger\partial_{LL}\phi}{(1+\phi^\dagger\phi)^2}
\,.
\label{cp102chiralano}
\eeq

Much in the same way as in the Adler--Bell--Jackiw anomaly \cite{ABJ} and in \cite{NSVZ3}, the chiral current nonconservation
is exhausted by one loop in the {\em Wilsonian} sense. 
Invoking the superfield formalism, we can translate (\ref{cp102chiralano}) in the superfield language,
\beq
\partial_{RR}\mathcal{J}_{LL} = \frac 1{4\pi}\left[
D_L\frac{\partial_{RR} A\bar D_L A^\dagger}{(1+A^\dagger A)^2}-
\bar D_L\frac{\partial_{RR} A^\dagger \bar D_L A}{(1+A^\dagger A)^2}\right]
\,,
\label{50}
\eeq
where the right-hand side is {\em exact} in the Wilsonian sense.

Following the general arguments of \cite{KS,DS} it is convenient to rewrite
Eq. (\ref{50}) in a general form
\beq
\partial_{RR}\mathcal{J}_{LL}=-\frac12 D_L \mathcal{W}_R+\frac12 \bar {D}_L \bar{\mathcal{W}}_R\,,
\label{fis}
\eeq
where 
\beq
\mathcal{W}_R = -\frac 1 {2\pi} \frac{\partial_{RR} A \bar D_L A^\dagger}{(1+A^\dagger A)^2}\,.
\label{fisp}
\eeq 
The superfield $\mathcal{W}_R$ on the right-hand side was absent at the classical level. 
The expression for $\mathcal{W}_R$ contains $S_{LRR}$ as its lowest component \cite{KS},
namely,
\beq
\mathcal{W}_R = -S_{LRR}^\dagger +i\theta_R(T_{LLRR}+i\partial_{RR}j_{5,LL})+i\theta_R\theta_R^\dagger \partial_{LL}S_{LRR}^\dagger\,.
\eeq
Note that the coefficient in front of $i\theta_R$ contains the real and imaginary parts.
The former is the trace of the energy-momentum tensor, the latter is the divergence of the U(1) current.

With this information in hand we finally arrive at
\beqn
&&S_{LRR}=\frac 1{\sqrt{2}\,\pi}\frac{\partial_{RR}\phi^\dagger \psi_L}{(1+\phi^\dagger\phi)^2}\,,\nonumber\\[3mm]
&&T_{LLRR}=-\frac{1}{2\pi}\left[
\frac{\partial_{LL}\phi^\dagger\partial_{RR} \phi+\partial_{RR}\phi^\dagger\partial_{LL}\phi}{(1+\phi^\dagger\phi)^2}
+\frac{2\psi_L^\dagger i\mathcal{D}_R \psi_L}{(1+\phi^\dagger\phi)^2}
\right]\,.
\label{54}
\eeqn

The first line in (\ref{54}) presents the superconformal anomaly while the second line presents the scale anomaly.
We see that in the Wilsonian sense these anomalies are exhausted by one loop. In particular, for the
trace of the energy-momentum tensor we obtain
\beq
T_{LLRR} =-T_\mu^\mu= \frac{\beta(g^2)}{g^2}\mathcal{L}_A\,.
\label{55}
\eeq 
Comparing the right-hand sides of (\ref{54}) and (\ref{55}) we conclude that
\beq
\beta_{\rm Wilsonian} = -\frac{g^4}{2\pi}\,.
\eeq
The denominator in Eq. (\ref{39}) is of an infrared origin. One can say that it comes at the stage of 
taking the matrix element of the right-hand side of (\ref{54}). Alternatively, one  can say that
it appears in passing from the holomorphic coupling to the canonically normalized coupling \cite{AH}.
One should compare this with exactly the same situation in four-dimensional supersymmetric gluodynamics \cite{NSVZ2,NSVZ3,AH}.

\section{Adding fermions}
\label{sec:2}

In this section we consider a more general (nonminimal) version of the $\mathcal{N}=(0,2)$ $\mathbf{CP}(1)$ nonlinear model, 
which is analogous to four-dimensional $\mathcal{N}=1$ super-Yang--Mills theory with adjoint matter. 
This will turn out beneficial for two reasons: first, we will strengthen the case for our all-loop exact $\beta$ function. 
Second, we will find a conformal window in multiflavor heterotic $\mathbf{CP}(1)$ models.

To add ``matter" we have to use the $\mathcal{N}=(0,2)$ superfield $B_i$ ($i=1,2, ..., N_f$) (see e.g.
\cite{CS1})
with the following structure
\beq
B_i(x, \theta_R, \theta_R^\dagger) = \psi_{R,i}(x)+\sqrt{2}\theta_R F_i(x)+i\theta_R^\dagger\theta_R\partial_{LL}\psi_{R,i}(x)\,.
\label{defBifield}
\eeq
As usual, the $F$ terms are auxiliary. Thus, the superfield $B$ contains only a single right-handed fermion degree of freedom (per flavor).
The latter has no bosonic counterpart.
Then the heterotic $\mathbf{CP}(1)$ model with matter acquires  the following
Lagrangian:
\beq
\mathcal{L} =\mathcal{L}_A+\mathcal{L}_B=  \int d^2 \theta_R \left[
\frac 1{g^2} \frac{A^\dagger i\overset{\leftrightarrow}{\partial_{RR}} A}{1+A^\dagger A}+
\frac 12\frac{\vec{B}^\dagger\cdot \vec B }{(1+A^\dagger A)^2}\right]\,, 
\label{58}
\eeq
where $\vec B$ is the vector made of fermionic chiral superfields,
\beq
\vec B = \left\{ B_i(x, \theta_R, \theta_R^\dagger)\right\}\,.
\label{defBifieldp}
\eeq

 It is easy to see that if $N_f=1$, the model (\ref{58})  reduces to the $\mathcal{N}=(2,2)$ $\mathbf{CP}(1)$ model.
 This circumstance will be used below.
The $ B_i$ fields live on the tangent space of $\mathbf{CP}(1)$, and,
 hence, are endowed with the following target space symmetry transformation:
\beq
\vec B \to \vec B + 2\bar \epsilon A\vec B\,,\quad \vec B^\dagger\to \vec B^\dagger + 2\epsilon A^\dagger \vec B^\dagger\,.
\eeq
 We will find an analog of the NSVZ $\beta$ function
 which replaces that of the minimal model (see Sec.~\ref{sec:1}).
 To this end we will exploit
 (a) instanton calculus and (b) hypercurrent analysis. We recall that adding fermions in the way described above is only possible for 
 $\mathbf{CP}(1)$. The only exception is the case $N_f=1$. In this case we deal with the nonchiral $\mathbf{CP}(1)$ model which can be readily generalized to
 $\mathbf{CP}(N-1)$.
 
 \subsection{Two-loop result: direct calculation}
 
\begin{figure}
 \begin{center}
 \includegraphics[width=3in]{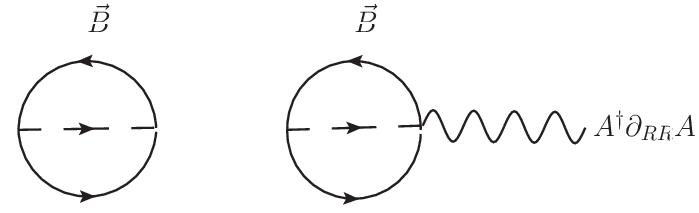}
 \end{center}
 \caption{\footnotesize Two-loop correction to $\beta(g^2)$ due to the  $B_i$ loops.}
 \label{super2loopf_111031}
 \end{figure}

First we will show that at the perturbative calculation at two-loop level gives exactly the answer we
 expect. We will collect two-loop corrections to the renormalization of $g^{-2}$
  by considering the quantum correction to the kinetic term of the
  superfield $A$. As compared with   the previous case, all we have to change is that now  we 
   need to take the $B_i$ loop into account, see Fig.~\ref{super2loopf_111031}. 
 
For each flavor  there will be a single diagram that contributes.
At the two-loop level distinct flavors do not interfere with each other, they enter additively. Hence,
we  can guess the result from what happens in the $\mathcal{N}=(2,2)$ $\mathbf{CP}(1)$ model, in which case the single-pole contribution due to $B$
cancels  that due to the $A$ loop. 
Indeed, as was mentioned above, if $N_f=1$
we deal with the $\mathcal{N}=(2,2)$ model, in which all loops in the charge renormalization higher than the first loop must cancel.

So without actually having to do the calculation (which is not difficult, though), we can write down the final result
\beq
\delta\mathcal{L}_{B\,{\rm two\,\,loop}} = \int d^2\theta_R N_f \frac{g^2}2\frac{i}{4\pi}  \frac{A^\dagger i\partial_{RR} A}{1+A^\dagger A} I+\text{H.c.}\,,
\eeq
where $I$ is defined in Eq.~(\ref{8}).
As a result we have, at the two-loop level
\beq
\beta(g^2)\equiv\frac{\partial}{\partial \text{log}\,M}g^2
=-\frac{g^4}{2\pi}\left(1-\frac{N_f-1}{4\pi} g^2\right)\,,
\label{62}
\eeq
where $M$ is the mass of the ultraviolet regulator (e.g. the Pauli--Villars mass).
Shortly we will see that Eq. (\ref{62}) can be rewritten as 
\beqn
\beta(g^2)&=&-\frac{g^4}{2\pi}\frac{1-\frac{N_f\,g^2}{4\pi} }{1 -\frac{g^2}{4\pi}}
\nonumber\\[3mm]
&=&
-\frac{g^4}{2\pi}\frac{1 +\frac{N_f\gamma}{2} }{1-\frac{g^2}{4\pi}}\,,
\label{63}
\eeqn
where $\gamma$ is the anomalous dimension\,\footnote{The anomalous dimension is defined below in (\ref{65}).}
 of the $B$ fields (which is one and the same for all matter fields due to the flavor symmetry
of the model (\ref{58})). Needless to say, at $N_f=1$ the $\beta$ function degenerates into
a one-loop expression. This is welcome since at $N_f=1$ we, in fact, deal with the ${\mathcal N}=(2,2)$ model,
whose $\beta$ function is exhausted by one loop \cite{Morozov:1984ad}.
 At the two-loop level Eqs. (\ref{62}) and (\ref{63}) are identical. In higher orders
(\ref{63}) is exact, as we will argue below.

How does the anomalous dimension $\gamma$ appear?
In 
the Lagrangian (\ref{58}), as we evolve it from $M$ down to a current normalization point $\mu$,
  we should take care of the  wave-function renormalization of the $B_i$ fields, in addition to the
  $g^{_2}$ renormalization. The fields $B_i$ live on the tangent space of the target manifold, and the covariant structure 
is uniquely fixed. 
\beq
\mathcal{L}_{B,\,UV} = \int d^2\theta_R\frac12\frac{\vec B_0^\dagger \vec B_0}{(1+A^\dagger A)^2}\,,\quad
\mathcal{L}_{B,\,IR} = \int d^2\theta_R \frac Z2\, \frac{\vec B_0^\dagger \vec B_0}{(1+A^\dagger A)^2}\equiv \int d^2\theta_R\frac 12\frac{\vec B_r^\dagger \vec B_r}{(1+A^\dagger A)^2}\,.
\eeq
 The $Z$ factor can be absorbed into the redefinition of $B_i$. It leaves a remnant, however, in the form of the
 Konishi anomaly, in much  the same way as in four-dimensional super-Yang--Mills \cite{NSVZ3}.
 
Alternatively, we could introduce a $Z$-factor in the ultraviolet as follows. We denote it by $Z_0$,
\beq
\mathcal{L}_{B,\,UV} =\int d^2\theta_R\frac12 Z_0\frac{\vec B^\dagger \vec B}{(1+A^\dagger A)^2}\,,\quad
\mathcal{L}_{B,\,IR} = \int d^2\theta_R\frac12\frac{\vec B^\dagger \vec B}{(1+A^\dagger A)^2}\,.
\eeq
Note that $Z_0 = Z^{-1}$.
These two renormalization schemes are consistent. The ultraviolet factor $Z_0$ is used in instanton calculus, see Sec.~\ref{instsubsec}. We introduce it here to make easier the comparison with the previous instanton calculations, for example \cite{NSVZ2, Morozov:1984ad, miar}

For each flavor, we have one and the  same diagram for the $Z$ factor which, after a simple and straightforward calculation, yields
\beq
Z = \frac1{Z_0} \equiv \left(\frac{B_{i,r}}{B_{i,0}}\right)^2 = 1-ig^2I\,,
\label{defZfactor}
\eeq
and
\beq
\gamma(\vec B) \equiv - \frac{\partial}{\partial\text{ log }\mu}\text{ log }Z \equiv  - \frac{\partial}{\partial\text{ log }M}\text{ log }Z_0= -\frac{g^2}{2\pi}+O(g^6)\,.
\label{65}
\eeq

\subsection{Instanton calculus}
\label{instsubsec}

In this section we will  apply the instanton analysis in the multiflavor model to substantiate 
Eq. (\ref{63}). In fact, the difference between the $N_f$-flavor model and the minimal model in essence reduces
to a different number of the fermion zero modes in the given 
one-instanton background. The target space symmetry 
 ensures that there are zero modes associated with the right-handed fermions $\vec{\psi_z}$. There are two of those
 for each ``matter" field,
\beq
\psi_{z,i}=\frac {y\alpha_i^\dagger}{ (z- z_0)^2}\,,\quad \psi_{z,i}=\frac{y\beta_i}{z-z_0}\,.
\eeq
Note that in the instanton measure we no longer have the factor ${g^2}$ for each pair of fermion zero modes of $\vec \psi_z$.
This is in contradistinction with what we had for the modes 
of the fermion component of the superfield $A$. The 
 normalization of the $B$ terms in (\ref{58}) is  canonical.

At the loop level we get corrections  to the collective coordinates $\alpha_i^\dagger$ and $\beta_i$ due to
the corresponding wave-function renormalization. Each fermion superfield acquires $\sqrt{Z}$, and so do $\alpha_i^\dagger$ and $\beta_i$.
Correspondingly, each $d\alpha_i^\dagger$ and $d\beta_i$
 will introduce a factor $Z^{-\frac 12}$, where $Z$ is defined as in (\ref{defZfactor}) and (\ref{65}). Therefore, summarizing, for each ``matter" fermion field, we have the accompanying
 factor $(ZM)^{-1}$.

The second question we must address is the one-loop correction due to nonzero modes. 
Following the same road as in Sec.~\ref{sec:inst}, we can build the  expansion in the nonzero modes using the eigenfunctions $u_n$ and $v_n$ defined in (\ref{fnzmodeseqn}).
Indeed, each flavor will give
\beq
\int [\mathscr{D}\delta\psi_{z,i}][\mathscr{D}\delta\psi_{\bar z,i}^\dagger]\rightarrow \left\{\det \left[\begin{array}{cc}0 & i\partial_z\frac 1{\chi_{\rm inst}^2} \\i\frac 1{\chi_{\rm inst}^2}\partial_{\bar z} & 0\end{array}\right] \right\}^{\frac12}=\left(\prod_n E_n^2\right)^{\frac12}\,,
\eeq
and, hence, each extra flavor will contribute $M$ in the instanton measure after an appropriate regularization of the infinite product. One can easily see that when $N_f=1$, we recover the fact that the one-loop determinant from the boson and fermion nonzero modes, respectively, cancel each other. 
This is certainly what we expect \cite{Morozov:1984ad,NSVZ}
As a result,  at the 
end of the day, we get the following expression for the measure:
\beqn
d\mu &=&\left( \frac{ M^2}{g_0^2} \right)^2\left( \frac{g_0^2}{M} \right)^{1} \left(\frac 1{Z_0M}\right)^{N_f} M^{-1+N_f} e^{-\frac{4\pi}{g_0^2}} \nonumber\\[2mm]
&\times & d\text{log}(y) d\text{log}(\bar y) \,\, dz_0 d\bar z_0 \,\, d\alpha d\beta^\dagger\prod_{i=1}^{N_f} d\alpha_i^\dagger d\beta_i
\,.
\label{67}
\eeqn

\begin{figure}
\begin{center}
\includegraphics[width=2.5in]{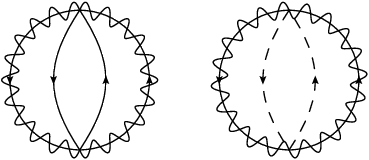}
\end{center}
\caption{\footnotesize An illustration of  how the cancellation at higher-loop level happens. The dashed lines 
are the $\phi$ propagators, the solid lines are those of $\psi_{\bar z}$, and the solid lines with the wavy lines superimposed denote the propagators of $\psi_{z,i}$.}
\label{super3loop_111123}
\end{figure}

Next, we note that 
our  nonrenormalization theorem in the instanton background 
derived in the minimal model (Sect.~\ref{sec:nsvz})  holds in the nonminimal model too. 
The general argument telling us that in the instanton background, all one-particle irreducible diagrams with two loops or more
do not contribute  is essentially the same as in Sect.~\ref{sec:nsvz}.
We can illustrate how it happens in the component language for three-loop graphs  shown 
in Fig.~\ref{super3loop_111123}.
The diagrams displayed on the left and on the right cancel each other.

Recall that the $Z$ factors 
of the $B_i$ fields get renormalized. 
These are one-particle reducible graphs in the instanton background  not   seen in the above
consideration (in the instanton background the $\psi_{z,i} $ kinetic terms vanish due to equations of motion). 
They have to be included in the measure additionally, as was done in (\ref{67}).

Asserting that the overall dependence of 
the instanton measure $d\mu$ on the ultraviolet cut-off $M$ should cancel, we arrive at the exact relation between 
the $\beta$ function and the anomalous dimension $\gamma(B_i)$,
\beq
\beta(g^2) = -\frac{g^4}{2\pi}\frac {1+\frac{N_f}2 \gamma(B_i)}{1-\frac{1}{4\pi}g^2}
\,,
\label{68}
\eeq
exactly as in (\ref{63}).

In the multiflavor model  neither $\beta(g^2)$ nor $\gamma(B_i)$ are all-loop exact. But the relation between
them is exact. This is similar to the situation in $\mathcal{N}=1$ super-Yang--Mills theory with   matter in four dimensions. As in the NSVZ $\beta$ function, 
the knowledge of $Z$'s  at one-loop order  gives $\beta(g^2)$ at two-loop order, and so on. 

\section{Hypercurrent for \boldmath{$N_f$} flavors}
\label{hcnf}

Now we can generalize the hypercurrent, passing from the minimal model (Sect. \ref{sec:current})
to the multiflavor model. At the
classical level the operator $\mathcal{J}_{LL}$ is defined exactly 
in the same way as in the minimal $\mathcal{N}=(0,2)$ model.  

\begin{figure}
\begin{center}
\includegraphics[width=2in]{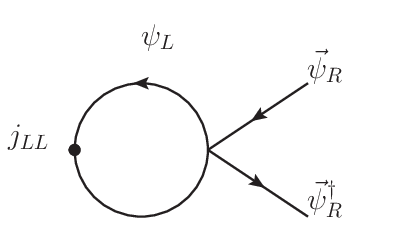}
\end{center}
\caption{\footnotesize One-loop diagram for the
$j_{LL}$ anomaly in the $\mathcal{N}=(0,2)$ $\mathbf{CP}(1)$ models with matter.}
\label{j2ano_111031b}
\end{figure}

The current $j_{LL}$ is corrected at the quantum level through the 
anomalous diagram depicted in Fig.~\ref{j2ano_111031a} and, in addition, through
 a new diagram shown in Fig.~\ref{j2ano_111031b}.
Now the anomaly can be expressed in superfields as
\beq
\partial_{RR}\mathcal{J}_{LL} = \frac 1{4\pi}\left[
D_L\frac{\partial_{RR} A\bar D_L A^\dagger}{(1+A^\dagger A)^2}-
\bar D_L\frac{\partial_{RR} A^\dagger D_L A}{(1+A^\dagger A)^2}+
\frac{g^2}2\frac{\{D_L,\bar D_L\}}{2i} \frac{2\vec B^\dagger \vec B}{(1+A^\dagger A)^2}\right]
\,.
\label{69}
\eeq
It is not difficult to understand that the last term on the right-hand side is just the leading term of expansion of the
exact (Wilsonian) formula,
\beq
\partial_{RR}\mathcal{J}_{LL} = \frac 1{4\pi}\left[
D_L\frac{\partial_{RR} A\bar D_L A^\dagger}{(1+A^\dagger A)^2}-
\bar D_L\frac{\partial_{RR} A^\dagger D_L A}{(1+A^\dagger A)^2}\right] -
\frac{1}{4}\, \gamma \, \frac{\{D_L,\bar D_L\}}{2i} \frac{2\vec B^\dagger \vec B}{(1+A^\dagger A)^2} 
\,.
\label{70}
\eeq
To substantiate this point let us consider the renormalization-group evolution of the bare Lagrangian (\ref{58}). 
The exact Wilsonian effective Lagrangian has the form
\beq
\mathcal{L}_{\rm Wilsonian} =  \int d^2 \theta_R \left[
\frac i{2g^2_h} \frac{A^\dagger \partial_{RR} A}{1+A^\dagger A}
-\frac i{2\bar{g}^2_h} \frac{A \partial_{RR} A^\dagger}{1+A^\dagger A}+
\frac {1}{2}\,\frac{\vec{B}_r^\dagger\cdot \vec B_r }{(1+A^\dagger A)^2}\right]\,, 
\label{71}
\eeq
where $g_h^2$ stands for the holomorphic running coupling whose renormalization is exhausted by one loop.
The response of this Lagrangian to scale transformations reduces to
\beq
\delta \mathcal{L}_{\rm Wilsonian} \propto  
\int d^2 \theta_R \left[
-\frac {i\beta_{\rm Wilsonian}}{2g^4_h} \,\,\frac{A^\dagger \partial_{RR} A}{1+A^\dagger A}
+\frac {i\bar{\beta}_{\rm Wilsonian}}{2\bar{g}^4_h} \,\,\frac{A\partial_{RR} A^\dagger}{1+A^\dagger A}
-\frac {\gamma}{2}\,\frac{\vec{B}_r^\dagger\cdot \vec B_r }{(1+A^\dagger A)^2}\right]\,.
\label{72p}
\eeq
The expression inside the square brackets in (\ref{72p}), up to a minus sign, is just another component of (\ref{51}). Thus, Eq. (\ref{72p}) confirms (\ref{70}). In components
Eq. (\ref{70}) is equivalent to 
\beqn
S_{LRR}&=&\frac 1{\sqrt{2}\pi}\frac{\partial_{RR}\phi^\dagger \psi_L}{(1+\phi^\dagger\phi)^2}\,,\nonumber\\[3mm]
T_{LLRR}&=&-\frac{1}{2\pi}\left[
\frac{\partial_{LL}\phi^\dagger\partial_{RR} \phi+\partial_{RR}\phi^\dagger\partial_{LL}\phi}{(1+\phi^\dagger\phi)^2}
+\frac{2\psi_L^\dagger i\mathcal{D}_{RR} \psi_L}{(1+\phi^\dagger\phi)^2}\right]\nonumber\\[3mm]
&&+\gamma\,\left[\frac{\vec{\psi}_R^\dagger i\overset{\leftrightarrow}{\mathcal{D}}_{LL} \vec{\psi}_R}{(1+\phi^\dagger\phi)^2}-\frac{2\psi_L^\dagger\psi_L\vec{\psi}_R^\dagger\vec{\psi}_R}{(1+\phi^\dagger\phi)^4}\right]\,.
\eeqn

Returning to Eq. (\ref{70}) we observe that the last term on the right-hand side
will convert itself into the first term through the Konishi anomaly (Sect. \ref{konishi}). This will produce the numerator of the $\beta$
function, cf. Eq. (\ref{63}). The denominator will appear upon taking the matrix element of the
operator $A^\dagger \partial A/(1+ A^\dagger A)$, or, alternatively, upon the transition from the
holomorphic coupling to the canonical coupling.

\section{``Konishi" anomaly}
\label{konishi}

As was mentioned in the Introduction,
 adding ``flavor" fields $B_i$ in the tangent space is similar to adding adjoint matter in the $\mathcal{N}=1$ four-dimensional 
 super-Yang--Mills theory. This similarity extends rather far. In particular, in this section we will derive an analog of the 
 Konishi anomaly  \cite{kenk}.

For each   matter field that we introduced, we  have an extra (classical) U(1) symmetry, 
corresponding to individual rotations of the $B_i$ fields, see Eq. (\ref{58}). It is obvious that the corresponding classically
conserved U(1) currents are
\beq
j_{RR, i} = \frac{g^2}2G\psi_{R,i}^\dagger\psi_{R,i}\,.
\label{74}
\eeq
These currents are
the lowest components of the superfield operators
\beq
 \mathcal{J}_{RR,i} \equiv  \frac 1{(1+A^\dagger A)^2} B_i^\dagger B_i\,,\qquad i= 1,2, ..., N_f\,.
\eeq

\begin{figure}
\begin{center}
\includegraphics[width=4in]{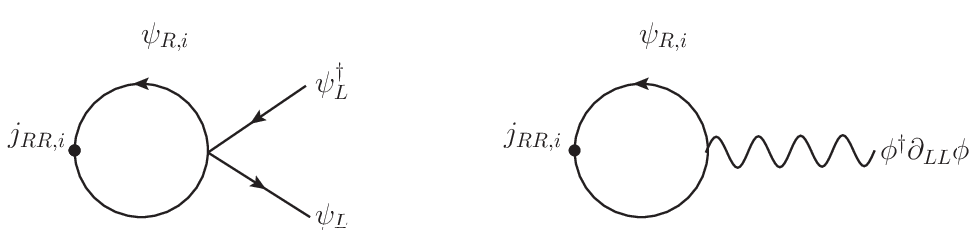}
\end{center}
\caption{\footnotesize One-loop correction to the $U(1)$ current $j_{RR,i}$.}
\label{j2ano_111031c}
\end{figure}

At the quantum level, due to the anomaly, these matter U(1) currents cease to be conserved. Instead,
  evaluating the diagrams in Fig.~\ref{j2ano_111031c}, we find
\beq
\partial_{LL} \mathcal{J}_{RR,i} = -\frac 1{4\pi}\left[
D_L\frac{\partial_{RR} A\bar D_L A^\dagger}{(1+A^\dagger A)^2}-
\bar D_L\frac{\partial_{RR} A^\dagger  D_L A}{(1+A^\dagger A)^2}\right]
\,,
\label{76}
\eeq
 for each (fixed) value of $i$. 
 Comparing this expression with the last term in Eq. (\ref{70})
(in its imaginary part)
we see that Eq. (\ref{70})
can be rewritten as follows:
\beq
\partial_{RR}\mathcal{J}_{LL} = \frac {1+ (N_f\gamma/2)}{4\pi}\left[
D_L\frac{\partial_{RR} A\bar D_L A^\dagger}{(1+A^\dagger A)^2}-
\bar D_L\frac{\partial_{RR} A^\dagger D_L A}{(1+A^\dagger A)^2}\right] 
\,.
\label{77}
\eeq
The real part of the superfield relation   takes the form 
\beq
T_{LLRR}= -\frac{1+(N_f\gamma/2)}{2\pi}\left[
\frac{\partial_{LL}\phi^\dagger\partial_{RR} \phi+\partial_{RR}\phi^\dagger\partial_{LL}\phi}{(1+\phi^\dagger\phi)^2}
+\frac{2\psi_L^\dagger i\mathcal{D}_{RR} \psi_L}{(1+\phi^\dagger\phi)^2}\right].
\eeq
These are still   Wilsonian operator formulas which present a direct parallel with, say, Eq. (2.111) in \cite{miar}.

It is clear that the passage to the generator of the 1-particle irreducible vertices (or, alternatively, from the holomorphic coupling to the canonic coupling \cite{AH}) proceeds exactly in the same manner as in the minimal model, resulting in the replacement
\beq
1+(N_f\gamma/2) \to \frac{1+(N_f\gamma/2)}{1- g^2/4\pi}\,.
\eeq

\section{Conformal window}
\label{conwid}

The $\beta$ function
that we have just derived, see Eq. (\ref{63}), has a remarkable property. Assume that $N_f\gg 1$. Then it develops an infrared fixed point
at parametrically small values of $g^2$, such that one can still trust the one-loop result for the
anomalous dimension of the matter fields,
\beq
\frac{g^2_*}{2\pi} = \frac{2}{N_f} \ll 1\,,
\eeq
where the asterisk labels the fixed point. If we choose the bare coupling constant in the
interval $(0, 2/N_f)$ then the theory under consideration is
asymptotically free in  the ultraviolet and conformal in the infrared. At large $N_f$
it is weakly coupled at all distances. Perturbative calculations of the anomalous dimensions of various operators make sense.

It is clear that on the side of small $N_f$
the conformal window should extend to some $N_f^* > 1$. At $N_f=1$ supersymmetry of the model under consideration is enhanced up to (2,2), and this model is certainly nonconformal. Rather, it develops a mass gap.

\section{Conclusion}
\label{conclu}

In this paper we studied a class of two-dimensional $\mathcal{N}=(0,2)$ nonlinear sigma models with $\mathbf{CP}(1)$ as the target space. We presented a number of arguments indicating the similarity of these models
 with four-dimensional super-Yang--Mills with adjoint matter. Two further questions could be asked here.
 
\begin{itemize}
\item Can we generalize our nonrenormalization theorem to other   models?
\item Can we see further implications of the 2D/4D correspondence?
\end{itemize}

These two questions are interrelated. As for the first one,   the answer is positive, at least in part. The original NSVZ argument can be applied to a large class of models with flag manifolds as 
the target manifolds. We do expect our analysis to go  through. 

The fermion anomaly \cite{MN} does not allow us to
extend our multiflavor models in the direction of $\mathbf{CP}(N-1)$. This is due to the fact that
the anomaly free condition $p_1=0$ ($p_1$ is the first Pontryagin class)   rules out  $\mathbf{CP}(N-1)$ target spaces except for $N_f=1$. If $N_f=1$, the model becomes nonchiral.

There are two obvious ``technical" questions to be explored. First, the (bi)fermion condensates.
It is well-known that such a condensate develops \cite{NSVZ}
in the ${\mathcal N}=(2,2)$ $\mathbf{CP}(1)$ model. Moreover, it plays the role of the order parameter distinguishing between two distinct vacua of this theory.
In the minimal ${\mathcal N}=(0,2)$ $\mathbf{CP}(1)$ model such a phenomenon seems to be impossible since
it is impossible to build a Lorentz scalar from the $\psi_L$ field alone.
We checked that one instanton in the minimal model does not give rise to the fermion condensate.
Whether or not they develop at $N_f>2$ remains to be seen.

Next, we will explore \cite{CS4} the NSVZ-like $\beta$ functions derived in this paper
in models including additional chiral fields, such as \cite{CS1}. At the moment we can formulate a conjecture that
the relation (\ref{68}) will stay intact; all dependence on the additional
coupling constants will be hidden in the anomalous dimensions.

Another issue of interest is the occurrence of conformality.
The nonsupersymmetric (bosonic)  $\mathbf{CP}(1)$ model is known to be conformal when the vacuum angle $\omega$ (see Eq.~(\ref{holomorcoup})) equals to $\pi$. Is it a
hint  that
the conformal window of the  ${\mathcal N}=(0,2)$ models  extends all the way down
to $N_f=2$?

\section*{Acknowledgments}

We thank  J. Chen, A. Vainshtein and J. Yagi
for illuminating discussions. XC is grateful to the 2011 Simons workshop in
Mathematics and Physics and the Simons Center for Geometry and Physics for hospitality, where part of the work was done.

The work of XC and MS was supported in part by the DOE grant DE-FG02-94ER40823.


\section*{Appendix A: Notations in Minkowski spacetime}
\renewcommand{\theequation}{A.\arabic{equation}}
\setcounter{equation}{0}

In this appendix we give a description of $\mathcal{N}=(0,2)$ $D=1+1$ superspace and fix the notations. 

The space-time coordinate $x^\mu=\{t,z\}$ can be promoted to superspace by adding a complex Grassmann variable $\theta_R$ and its complex conjugate $\theta_R^\dagger$. Wherever our expressions are dependent on the representation of Clifford algebra, we use the following convention.
\beq
\gamma^0=\left(\begin{array}{cc} 0 & -i\\ i & 0\end{array}\right)\,,\quad \gamma^1=\left(\begin{array}{cc} 0 & i\\ i & 0\end{array}\right)\,, \quad\gamma^3=\gamma^0\gamma^1\,.
\eeq
Under this representation the Dirac fermion is expressed as 
\beq
\psi=\left(\begin{array}{c}\psi_R\\ \psi_L\end{array}\right)\,.
\eeq 

We define the left moving and right moving derivatives as
\beq
\partial_{LL}=\partial_t+\partial_z\,,\qquad \partial_{RR}=\partial_t-\partial_z\,,
\label{72}
\eeq 
and use the following definition for the superderivatives:
\beq
D_L = \frac \partial {\partial \theta_R}-i\theta_R^\dagger \partial_{LL}\,,\qquad
\bar D_L = -\frac\partial{\partial \theta_R^\dagger}+i\theta_R\partial_{LL}\,.
\eeq
Their commutator gives $\{D_L,\bar D_L\} = 2i\partial_{LL}\,$.

To change between ordinary coordinates and the lightcone coordinates, we have, for supercurrent:
\beq
S^0_L = S_{RRL}+S_{LLL}\,,\quad S^1_L=S_{LLL}-S_{RRL}\,.
\eeq
And for $T^{\mu\nu}$:
\beqn
T_{LLLL}&=T_{00}+T_{10}+T_{11}+T_{01}\,,\\
T_{LLRR}&=T_{00}+T_{10}-T_{11}-T_{01}\,,\\
T_{RRLL}&=T_{00}-T_{10}-T_{11}+T_{01}\,,\\
T_{RRRR}&=T_{00}-T_{10}+T_{11}-T_{01}\,.
\eeqn

The shifted space-time coordinates that satisfy the chiral condition are 
\beq
y^0=t+i\theta_R^\dagger\theta_R\,,\qquad y^1=z+i\theta_R^\dagger\theta_R\,.
\label{defy}
\eeq 
The antichiral counterparts are
\beq
\tilde y^0=t-i\theta_R^\dagger\theta_R\,,\qquad \tilde y^1=z-i\theta_R^\dagger\theta_R\,.
\label{defyt}
\eeq 
Under supersymmetric transformation $ \delta_\epsilon+\delta_{\bar \epsilon}$
\beqn
 \theta_R \to \theta_R+ \epsilon\,,&\qquad&  \theta^\dagger_R \to \theta^\dagger_R+ \bar\epsilon\,,\nonumber\\[3mm]
 y^\mu \to y^\mu+ 2i\bar\epsilon\theta_R\,,&\qquad&  \tilde y^\mu \to \tilde y^\mu -2i\theta^\dagger_R\epsilon\,,
\eeqn
where $\mu = 0,1$.

We can now define the chiral $\mathcal{N}=(0,2)$ bosonic and fermionic superfields in our model,
\beqn
&&
A(y^\mu,\theta_R)=\phi(y^\mu)+\sqrt{2}\theta_R\psi_L(y^\mu)\,,\nonumber\\[3mm]
&&B(y^\mu,\theta_R)=\psi_R(y^\mu)+\sqrt{2}\theta_RF(y^\mu)\,.
\label{fieldsdef}
\eeqn
Here $\phi$, $\psi_L$ and $\psi_R$ describe physical degrees of freedom, while $F$ will enter without derivatives and, thus, can be eliminated by virtue of equations of motion. 

\section*{Appendix B: Notations in Euclidean spacetime}

\renewcommand{\theequation}{B.\arabic{equation}}
\setcounter{equation}{0}

The Wick rotation is defined by
\beq
x^1 = x\,,\quad x^2 = it\,.
\eeq
Gamma matrices:
\beq
\gamma^1 = \left(\begin{array}{cc}0 & 1 \\1 & 0\end{array}\right)\,,\quad \gamma^2 = \left(\begin{array}{cc}0 & -i \\i & 0\end{array}\right)\,,\quad\gamma^3 = i\gamma^1\gamma^2\,.
\eeq
We define 
\beq
\partial_z = \partial_1-i\partial_2\,,\quad \partial_{\bar z}=  \partial_1+i\partial_2\,,
\eeq
and the supercharges are given by
\beq
Q = \frac{\partial}{\partial \theta}+i\theta^\dagger \partial_{\bar z}\,,\quad \bar Q = -\frac{\partial}{\partial \theta^\dagger}-i\theta \partial_{\bar z}\,,
\eeq
together with the commutation relation
\beq
\{Q, \bar Q\} = -2i\partial_{\bar z}\,.
\eeq
Correspondingly, superderivatives are given by
\beq
D = \frac{\partial}{\partial \theta}-i\theta^\dagger \partial_{\bar z}\,,\quad \bar D = -\frac{\partial}{\partial \theta^\dagger}+i\theta \partial_{\bar z}\,.
\eeq
It is easy to verify that
\beq
D(z) = \bar D(z) = D(\bar z_{ch}) = 0\,,
\eeq
where $\bar z_{ch} = \bar z-2i\theta^\dagger\theta$.

We define the Dirac fermion to be
\beq
\psi = \left(\begin{array}{c}\psi_z \\\psi_{\bar z}\end{array}\right)\,.
\eeq

Now the bosonic chiral superfield is defined as
\beq
A = \phi(\bar z+2i\theta^\dagger\theta, z)+\sqrt{2}\theta\psi_{\bar z}(\bar z+2i\theta^\dagger\theta, z)\,.
\eeq
The fermionic chiral superfield can be written down in a similar manner.


\end{document}